\documentclass[12pt]{article}
\usepackage{graphicx,amssymb}
\usepackage{geometry}
\usepackage{array,dcolumn,hhline}
\usepackage[left]{lineno}
\renewcommand{\thefootnote}{\fnsymbol{footnote}}
\newcolumntype{d}{D{.}{.}{3.1}}
\newcolumntype{p}{D{.}{.}{4.2}}

\newcommand{\m}{\makebox[0.15mm][l]{ }}
\newcommand{\ma}{\makebox[16cm][c]}
\newcommand{\mb}{\makebox[6.5mm][l]{ }}

\newcommand{\mf}{\makebox[2.5mm][l]{ }}

\newcommand{\mi}{\makebox[4.0mm][l]}

\begin{document}
\setlength{\parindent}{0in}
\large
\ma{{\bf Will $^{37}$Ar emissions from light water power reactors become}} \\[3mm]
\ma{{\bf an obstacle to its use for nuclear explosion monitoring?}} \\[10mm]
\large
\ma{Gerald Kirchner$^{*}$, Franziska Gerfen, Anna Heise, Timo Schl\"uschen} \\
\small
\begin{center}
Universit\"at Hamburg, ZNF, Hamburg, Germany\footnotetext[1]{Corresponding author: Carl-Friedrich von Weizs\"acker-Centre for Science and Peace Research (ZNF), Universit\"at Hamburg, Beim Schlump 83, 20144 Hamburg, Germany; gerald.kirchner@uni-hamburg.de} \\[12mm]
\end{center}
\normalsize
\renewcommand{\thefootnote}{\arabic{footnote}}

         %
         %
\large
\ma{{\sf ABSTRACT}} \\[8mm]
\normalsize
$^{37}$Ar is a promising candidate for complementing radioxenon isotopes as indicators of underground nuclear explosions. This study evaluates its potential anthropogenic background caused by emissions from commercial pressurised water reactors. Various $^{37}$Ar production pathways, which result from activation of $^{36}$Ar and of $^{40}$Ca, respectively, are identified and their emissions quantified. In-core processes include (1) the 
restart of operation and degassing of the primary cooling water after maintenance and refueling shutdown, (2) the replacement of primary coolant water for limiting its tritium concentrations, and (3) the leakage of $^{37}$Ar produced from calcium impurities in UO$_{2}$ after fuel rod failures. Activation of air and of calcium in concrete within the biological shield are major out-of-core production pathways. Whereas emissions from in-core processes are transient, a rather constant $^{37}$Ar source term results from its out-of-core production. \\[2mm]
Generic atmospheric dispersion simulations indicate that already at moderate distances from the emitter, concentrations of $^{37}$Ar caused by routine reactor operations are far below its cosmogenic background in air. The only exception results from an inadvertent reactor re-start without operation of the primary cooling water degassing system for prolonged time. Such an event also causes high emissions of $^{41}$Ar which can be used for discriminating its $^{37}$Ar signal from an underground nuclear explosion. \\[10mm]
\large
\ma{{\sf KEYWORDS}} \\[8mm]
\normalsize
CTBT, treaty verification, $^{37}$Ar, anthropogenic background, PWR, $^{37}$Ar production routes \\[8mm]
\large
\ma{{\sf HIGHLIGHTS}} \\[3mm]
\normalsize
\begin{itemize}
\item Various $^{37}$Ar production pathways are identified in pressurised water reactors
\item Atmospheric emissions from these production pathways are quantified
\item Only an inadvertent reactor re-start without operating degassing system may result in atmospheric $^{37}$Ar concentrations at moderate distances in excess of its cosmogenic background
\item  $^{37}$Ar is a promising radioisotope for identifying underground nucler explosions
\end{itemize}
\m \\[8mm]
         %
         %
\large
{\sf 1. INTRODUCTION} \\[3mm]
\normalsize
The detection of radioisotopes created by nuclear explosions is a key component of the Comprehensive Nuclear-Test-Ban Treaty (CTBT) verification regime. For underground explosions the focus is on radioxenon due to its chemical inertness, high fission yield and its potential for discriminating radioxenon, which originates from a nuclear explosion, from reactor emissions if the four principal Xe radioisotopes are detected in atmospheric samples (Kalinowski and Pistner, 2006). However, intensified xenon monitoring by the International Monitoring System (IMS) of CTBT revealed that atmospheric emissions of radiopharmaceutical production facilities may create radioxenon signals with isotope ratios indistinguishable from those of nuclear explosions (Saey, 2009). Thus, including additional radionuclides might increase the sensitivity of the noble gas monitoring for CTBT verification. A promising candidate is $^{37}$Ar which is produced by activation of calcium during underground nuclear explosions and with a half-life of 35 days is still present in case of delayed emission into the atmosphere when much of the radioxenon has been decayed (Carrigan et al., 1996; Haas et al., 2010). Measurements of $^{37}$Ar are envisaged for CTBT on-site inspections (Aalseth et al., 2011), but the anthropogenic background of $^{37}$Ar, its major sources and their potential impact on nuclear explosion monitoring need to be assessed, before an extension of the IMS noble gas monitoring network should be recommended. However, this information is not readily available, as $^{37}$Ar atmospheric releases of nuclear facilities are not monitored. For argon, such data is available only for its short-lived isotope $^{41}$Ar (half-live 1.83 hours), which in nuclear reactors originates from the activation of $^{40}$Ar, the main natural argon isotope. Recently, $^{37}$Ar emissions have been measured at a research reactor (Johnson et al., 2017), but there is no information available for commercial nuclear power reactors. \\[2mm]
This study evaluates the production rates and release pathways of $^{37}$Ar in pressurised water reactors (PWR), which constitute about 66 \% of the operational nuclear power reactors (International Atomic Energy Agency, 2019). Measurements of $^{37}$Ar in reactor emissions, which have been performed for validating these evaluations, are presented elsewhere (Purtschert et al., in preparation). \\[9mm]
         %
         %
\large
{\sf 2. MATERIALS AND METHODS} \\[3mm]
\normalsize
{\bf 2.1 Reactor} \\[3mm]
All assessments are performed for the latest 1300 MW${\rm e}$ class of PWR reactors built by the German manufacturer Kraftwerk Union (KWU) which are still operating in Germany. They use UO$_{2}$ fuel with maximum burnup of 50 MWd per kg of heavy metal (HM). If not noted otherwise, the design information required for our analyses was taken from technical documents made available to the public during their licensing procedures. Even if the detailed design of modern PWR reactors built by other manufacturers may show some differences, the key results of this study are expected to apply for these also. \\[5mm]
{\bf 2.2 In core $^{37}$Ar production pathways} \\[3mm]
{\em 2.2.1 Pathway (1): Activation of argon in cooling water} \\[3mm]
During each refueling and maintenance shutdown, the reactor vessel is removed and the primary cooling loop is connected to the water-filled cooling pond inside the reactor building. Because these outages usually continue for some weeks, the concentration of air dissolved in the cooling water approaches equilibrium. For restart, the primary cooling loop is isolated and within approximately 1 day the power is increased stepwise to its nominal level (Fig. 1). When passing the reactor core, the argon dissolved in the primary cooling water becomes activated with $^{37}$Ar. It is produced by neutron capture of $^{36}$Ar, which is a natural argon isotope with 0.337 \% abundance. However, concentrations of dissolved gases and thus $^{37}$Ar production rates are efficiently reduced by the coolant purification system, which includes a vacuum evaporator for coolant degassing. In the KWU reactors, this system operates with a purification rate $L_{p}$ of 0.09 h$^{-1}$ which may be doubled if necessary. \\[2mm]
Thus, the dynamics of the numbers of $^{37}$Ar atoms, $N^{c}_{37}(t)$, and of its precursor $^{36}$Ar, $N^{c}_{36}(t)$, present in the primary coolant are given as
\begin{eqnarray}
    \frac{d N^{c}_{36}(t)}{d t} & = & -L_{p}  \cdot N^{c}_{36}(t) \: - \: p_{V} \cdot \displaystyle{\int_{E}} dE \: \sigma^{a}_{36} (E) \cdot \Phi (E) \cdot N^{c}_{36} (t) \\
    \frac{d N^{c}_{37}(t)}{d t} & = & - (\lambda_{37}+L_{p})  \cdot N^{c}_{37}(t) \: + \: p_{V} \cdot \displaystyle{\int_{E}} dE \: \sigma^{n,\gamma}_{36} (E) \cdot \Phi (E) \cdot N^{c}_{36} (t) \; \; ,
\end{eqnarray}
where $p_{V}$ is the fraction of the primary cooling water present within the active core zone (5.5 \% of 405 m$^{3}$), $\sigma^{a}_{36} (E)$, $\sigma^{n,\gamma}_{36} (E)$ denote the neutron absorption cross sections of $^{36}$Ar and its $^{36}$Ar$\,$(n,$\gamma$)$ \, ^{37}$Ar production cross section, respectively, and $\Phi(E)$ is the neutron flux density within the moderator volume of the fuel elements. With the initial $^{36}$Ar at restart, $N^{c}_{36}(t_{0})$, which follows from the mole fraction solubility of argon of 2.328$\cdot$10$^{-5}$ (International Union of Pure and Applied Chemistry, 1980) at 101.325 kPa and 30$^\circ$ C, the solution of eqns. (1), (2) for $t \ge t_{0}$ is given as
\begin{eqnarray}
  N^{c}_{37}(t) & = &  {\rm e}^{-(\lambda_{37}+L_{p}) \cdot (t-t_{0})} \nonumber \\[2mm]
  &  & \cdot \left[ N^{c}_{37}(t_{0})  + N^{c}_{36}(t_{0}) \, \frac{p_{V} \cdot \displaystyle{\int_{E}} dE \: \sigma^{n,\gamma}_{36} (E) \cdot \Phi (E)}{\lambda_{37} - p_{V} \cdot \displaystyle{\int_{E}} dE \: \sigma^{a}_{36} (E) \cdot \Phi (E)} \right. \\[2mm]
 & & \left. \makebox[25mm]{} \cdot \left( {\rm e}^{\displaystyle (\lambda - p_{V} \cdot \int_{E} dE \: \sigma^{a}_{36} (E) \cdot \Phi (E))\cdot (t-t_{0})} - 1 \right) \right] \nonumber ,
\end{eqnarray}
where $N^{c}_{37}(t_{0})$  denotes the residual $^{37}$Ar, which may be present in the primary cooling system from previous reactor operation. \\[2mm]
The highest $^{37}$Ar emissions from this pathway will occur if after a refueling shutdown the reactor is restarted, but the degassing system is started inadvertently only after some period of full power operation. \\[2mm]
Routinely, a small fraction (typically some m$^{3}$ d$^{-1}$) of primary cooling water is replaced and after purification, degassing and interim storage is discharged to avoid the accumulation of tritium. The makeup water is degassed before entering the cooling system. This process causes some accumulation of radioargon within the primary cooling system, as after the restart phase degassing is limited to the water fraction to be discharged. If we assume that (i) 10 m$^{3}$ of primary coolant are replaced daily, (ii) this exchange is a continuous process, and (iii) degassing reduces stable argon concentrations of the makeup water by 10$^{3}$, the accumulation of $^{37}$Ar in the coolant is given as
\begin{eqnarray}
   \frac{d N^{c}_{36}(t)}{d t} & = & N^{\rm in}_{36} \: - \:L_{r}  \cdot N^{c}_{36}(t) \: - \: p_{V} \cdot \displaystyle{\int_{E}} dE \: \sigma^{a}_{36} (E) \cdot \Phi (E) \cdot N^{c}_{36} (t) \\
    \frac{d N^{c}_{37}(t)}{d t} & = &  - (\lambda_{37}+L_{r})  \cdot N^{c}_{37}(t) \: + \: p_{V} \cdot \displaystyle{\int_{E}} dE \: \sigma^{n,\gamma}_{36} (E) \cdot \Phi (E) \cdot N^{c}_{36} (t) \; \; ,
\end{eqnarray}
where $N^{\rm in}_{36}$ denotes the input rate of $^{36}$Ar with the degassed makeup water and $L_{r}$ the cooling system's water replacement and degassing rate. At equilibrium, the number of $^{37}$Ar atoms present in the cooling water become 
\begin{eqnarray}
  N^{c,\rm eq}_{37} & = &   N^{\rm in}_{36} \: \cdot \: \frac{p_{V} \cdot \displaystyle{\int_{E}} dE \: \sigma^{n,\gamma}_{36} (E) \cdot \Phi (E)}{ ( L_{r} + p_{V} \cdot \displaystyle{\int_{E}} dE \: \sigma^{a}_{36} (E) \cdot \Phi (E) \, )  \, \cdot  \, ( L_{r} + \lambda_{37} ) }  \mf .
 \end{eqnarray}
{\em 2.2.2 Pathway (2): Activation of calcium in fuel} \\[3mm]
Due to its production process, UO$_{2}$ fuel shows minor calcium impurities with a mean value of 5 ppm (Andreas Schnieder, Framatome Advanced Nuclear Fuels, Personal communication). During irradiation, $^{37}$Ar is produced within the fuel matrix via the $^{40}$Ca\,(n,$\alpha$)\,$^{37}$Ar reaction. A fraction of this diffuses to the grain boundaries, into pores within or between grains and within connected pores into the fuel-cladding gap and gas plenum of the fuel rod (Rest et al., 2019). In the event of a fuel rod failure these gaseous radionuclides may be released into the primary cooling system. Since in a PWR such a failure today has a probability of about 10$^{-4}$ per rod (International Atomic Energy Agency, 2010), it is assumed that a single fuel rod will develop a leak at high burnup when it shows high fission gas concentrations in the fuel-cladding gap. \\[2mm]
For KWU fuel measurements have shown, that for operation regimes without major transients at 50 MWd (kg HM)$^{-1}$ burnup up to 10 \% of the fission gases have accumulated in the fuel-cladding gap and fission gas plenum (Hering, 1983; Manzel and Walker, 2002). In this gas, a fraction $p_{g}$ of 0.36 \% of the $^{37}$Ar fuel rod inventory will be present, if it is assumed that (i) the fuel release rate $f_{1}$ is time-invariant giving a value of about 9$\, \cdot \,$10$^{-10}$ s$^{-1}$, that (ii) argon behaves similar to the major fission gases krypton and xenon, and if (iii) the decay of $^{37}$Ar is taken into account. If a rod defect occurs, this radioargon can escape from the gap into the cooling water. Following Lewis et al. (2017), a gap escape rate $f_{2}$ of 2$\, \cdot \,$10$^{-5}$ s$^{-1}$ is a typical value for a large defect of a PWR fuel rod.  \\[2mm]
If a fuel rod defect develops at high burnup at time $t_{0}$, the numbers of $^{37}$Ar nuclei present in the fuel-cladding gap and the gas plenum of the affected rod, $N^{g}_{37}(t)$, and in the primary cooling water, $N^{c}_{37}(t)$,  are given as
\begin{eqnarray}
\frac{d N^{g}_{37}(t)}{d t} & = & - (\lambda_{37}+f_{2})  \cdot N^{g}_{37}(t) \: + \: f_{1} \cdot N^{f}_{37} \\[2mm]
\frac{d N^{c}_{37}(t)}{d t} & = & - (\lambda_{37}+L_{p})  \cdot N^{c}_{37}(t) \: + \: f_{2} \cdot N^{g}_{37}(t) \; ,
\end{eqnarray}
where $N^{f}_{37}$ is the $^{37}$Ar present in the fuel, which after prolonged operation at almost full power has attained equilibrium. With the initial conditions 
\begin{eqnarray}
 N^{g}_{37}(t_{0}) \, = \, p_{g} \cdot N^{f}_{37} & , & N^{c}_{37}(t_{0}) \, = \, 0
 \end{eqnarray}
  eq. (8) has the solution   
\begin{eqnarray}
  N^{c}_{37}(t) & = &  \frac{f_{2} \cdot p_{g} \cdot N^{f}_{37}}{L_{p} - f_{2}} \: \left( {\rm e}^{-(\lambda_{37}+f_{2}) \cdot (t-t_{0})} -  {\rm e}^{-(\lambda_{37}+L_{p}) \cdot (t-t_{0})} \right) \nonumber \\[2mm]
   & & + \frac{f_{1} \cdot f_{2} \cdot N^{f}_{37}}{(\lambda_{37} + f_{2}) \cdot (\lambda_{37} + L_{p})} \cdot \\[2mm]
   & & \cdot \left[ 1 + \frac{1}{L_{p} - f_{2}} \left( (\lambda_{37}+L_{p}) \cdot {\rm e}^{-(\lambda_{37}+f_{2}) \cdot (t-t_{0})} - (\lambda_{37}+f_{2}) \cdot {\rm e}^{-(\lambda_{37}+L_{p}) \cdot (t-t_{0})} \right) \right] \mb \nonumber
\end{eqnarray}
After a power transient, notably for reactor shutdown, enhanced releases of volatile fission products from defect fuel rods into the cooling water are observed (Lewis et al., 1997). For assessing the $^{37}$Ar emissions following such an event, it is assumed that a defect had developed at the top of the fuel rod shortly before reactor shutdown, which causes a spike release of the inventory of the gas plenum. The $^{37}$Ar present in the primary cooling system after such an event can be estimated as
\begin{eqnarray}
 N^{c}_{37}(t) & = & p_{g} \cdot N^{f}_{37} \cdot {\rm e}^{-(\lambda_{37}+L_{p}) \cdot (t-t_{0})} \; 
 \end{eqnarray} 
 if the small fuel-to-gap release rate of the fission gases is neglected. \\[5mm]
{\bf 2.3 Out of core $^{37}$Ar production pathways} \\[3mm]
{\em 2.3.1 Pathway (3): Activation of air} \\[3mm]
The reactor pressure vessel is enclosed by a massive concrete casing, the biological shield (Fig. 2), which attenuates the gamma and neutron radiation originating from the reactor core. It includes two cavities filled with air. The residual neutron flux causes the production of $^{37}$Ar by activation of the natural $^{36}$Ar isotope in both cavities. Since the outer cavity is used for cooling the concrete, it is part of a separated exhaust air system, which includes the rooms with contaminated air. It has a total volume of 10$^{4}$ m$^{3}$ with an air exchange rate $L_{o}$ of 10 \% h$^{-1}$ via filters into the main air exhaust system. The inner cavity of the biological shield is isolated, but shows a leak rate of 1 m$^{3}$ h$^{-1}$ into the outer cavity corresponding to an exchange rate $L_{i}$ of 0.625 \% h$^{-1}$. \\[3mm]
Since the neutron flux density does not vary radially within the cavities (see Fig. 6 below), $^{37}$Ar production rates are given as
\begin{eqnarray}
    \frac{d N^{i}_{37}(t)}{d t} & = & - (\lambda_{37}+L_{i})  \cdot N^{i,o}_{37}(t) \: + \: f_{i}\cdot N^{a}_{36} \cdot \displaystyle{\int_{E}} dE \: \sigma^{n,\gamma}_{36} (E) \cdot \Phi^{i} (E)  \\[2mm]
     \frac{d N^{o}_{37}(t)}{d t} & = & L_{i}  \cdot N^{i}_{37}(t) - (\lambda_{37}+L_{o})  \cdot N^{o}_{37}(t) \: + \: f_{o} \cdot N^{a}_{36} \cdot \displaystyle{\int_{E}} dE \: \sigma^{n,\gamma}_{36} (E) \cdot \Phi^{o} (E)    \; , \mi{}
\end{eqnarray}
where $i,o$ denote the inner and outer cavity, respectively, $N^{a}_{36}$ is the $^{36}$Ar concentration in natural air, and $f_{i,o}$ are the volume fractions parallel to the active core zone (0.43 for the inner and 6.42$\, \cdot \,$10$^{-3}$ for the outer cavity and its exhaust air system; see also Fig. 2). Assuming routine operation at rated power, equilibrium concentrations of
\begin{eqnarray}
    N^{i,{\rm eq}}_{37} & = & \frac{f_{i}\cdot N^{a}_{36}}{\lambda_{37}+L_{i}} \cdot \displaystyle{\int_{E}} dE \: \sigma^{n,\gamma}_{36} (E) \cdot \Phi^{i} (E) \\[2mm] 
    N^{o,{\rm eq}}_{37} & = & \frac{f_{o}\cdot N^{a}_{36}}{\lambda_{37}+L_{o}} \cdot \displaystyle{\int_{E}} dE \: \sigma^{n,\gamma}_{36} (E) \cdot \Phi^{o} (E) \nonumber \\[2mm]
    & & + \frac{L_{i} \cdot f_{i}\cdot N^{a}_{36}}{(\lambda_{37}+L_{o}) \cdot (\lambda_{37}+L_{i})} \cdot \displaystyle{\int_{E}} dE \: \sigma^{n,\gamma}_{36} (E) \cdot \Phi^{i} (E) 
\end{eqnarray}
are attained after about 12 days for the ventilation rates specified above. \\[3mm]
{\em 2.3.2 Pathway (4): Activation of calcium in concrete} \\[3mm]
The residual neutron flux produces $^{37}$Ar within the concrete of the biological shield via the $^{40}$Ca\,(n,$\alpha$)\,$^{37}$Ar reaction. The calcium content of concrete may vary widely between about 8.7 \%, if granite is used as aggregate, and 37.5 \% in case of limestone as aggregate (Gernot Kirchner, LafargeHolzim, Personal communication). The flux density, $j_{37}$, of the radioargon fraction, which enters the cavities of the biological shield, can be calculated by Fick's law as
\begin{eqnarray}
 j_{37}(r_{s},t) & = & - \epsilon \cdot D_{\rm eff} \cdot \left. {\frac{\partial \, C_{37}(r,t)}{\partial \, r}} \displaystyle \right|_{r=r_{s}}    \mi{} ,
\end{eqnarray}
where $r$ denotes the radial distance from the outer edge of the reactor core, $r_{s}$ is the cavity/concrete interface position, $\epsilon$ its porosity, $D_{\rm eff}$ denotes the $^{37}$Ar diffusion coefficient within the concrete and $C_{37}(r,t)$ the concentration of $^{37}$Ar produced via calcium activation. A generic value of the porosity $\epsilon$ of concrete is 2 \% (Gernot Kirchner, LafargeHolzim, Personal communication). The effective diffusion coefficient $D_{\rm eff}$ of a substance in a porous medium can be linked with its free molecular diffusion coefficient $D_{\rm m}$ by
\begin{eqnarray}
 D_{\rm eff} & = & \frac{\epsilon}{\tau} \cdot D_{\rm m}   \mi{} ,
\end{eqnarray}
where $\tau$ denotes the tortuosity of the pore space. It can be approximated by $\tau \, = \, \epsilon^{-1}$ (Shen and Chen, 2007). At standard conditions, $D_{\rm m}$ of argon in air has a value of 1.89$\, \cdot \,$10$^{-5}$ m$^{2}$ s$^{-1}$. With these data, the diffusion length
\begin{eqnarray}
 \ell_{37} & = & \sqrt{D_{\rm eff} \cdot \lambda^{-1}_{37}}
 \end{eqnarray}
of $^{37}$Ar in the concrete of the biological shield becomes 18.2 cm. \\[3mm]
After about 100 days of operation at rated power, $^{37}$Ar has attained its equilibrium concentration, $C^{\rm eq}_{37}(r)$, and eq. (16) gives a time-independent upper limit of its emissions from this pathway. Then, the radioargon concentration gradient in eq. (16) is given by the diffusion equation as
\begin{eqnarray}
 D_{\rm eff} \cdot  {\frac{\partial^{2} \, C_{37}^{\rm eq}(r)}{\partial \, r^{2}}} - \lambda_{37} \cdot  C_{37}^{\rm eq}(r) +  f \cdot  P_{37}(r) & = & 0 \mi{} ,
\end{eqnarray}
where $f$ is the $^{37}$Ar emanation fraction from the solid particles into the pore space, and $P_{37}(r)$ its volumetric production rate within the concrete
\begin{eqnarray}
    P_{37} (r) & = & N_{40} \cdot \displaystyle{\int_{E}} dE \: \sigma^{n,\alpha}_{40} (E) \cdot \Phi (r,E) \mi{} .
\end{eqnarray}
In eq. (20), $N_{40}$, $\sigma^{n,\alpha}_{40}$ denote the volumetric atom density of $^{40}$Ca in the concrete and its activation cross section, respectively. Since no data are available for the emanation fraction of argon, the value of 0.11$\, \pm \,$0.02 for the noble gas radon (Sahoo et al., 2011) has been adopted.  \\[3mm]
For evaluating eqs. (16), (19), the initial condition is set to $C_{37}(r,0)=0$. The following two cases are considered. \\[3mm]
{\em Case 1: Slab of thickness 2$\,$H} \\
If $H \lessapprox \ell_{37}$, the $^{37}$Ar profile within the slab is influenced by its diffusion through both surfaces. For a spatially constant production rate $P_{37}$ within the slab and the boundary conditions $C_{37}(\pm H) = 0$, we get (Sahoo et al., 2011)
\begin{eqnarray}
 j^{\rm eq}_{37} \, (\pm H) & = & f \cdot P_{37} \cdot \ell_{37} \cdot \tanh\left[\frac{H}{\ell_{37}} \right]    \mi{} .
\end{eqnarray}
{\em Case 2: Semi-infinite slab} \\
If $H \gtrapprox \ell_{37}$, eq. (18) approaches the well-known solution for a semi-infinite slab (e.g. D\"orr and M\"unnich, 1990)
\begin{eqnarray}
 j^{\rm eq}_{37} \, (\pm H) & = & f \cdot P_{37} \cdot \ell_{37}    \mi{} .
\end{eqnarray}
Fluxes of $^{37}$Ar into the cavities are obtained by multiplying eq. (21), (22) by the respective surface areas of the concrete's activated zone,which are derived from their radial distances from the center of the core (see Fig. 6) and the active length of the fuel elements of 3.9 m. \\[5mm]
{\bf 2.4 Noble gas retention systems and emission} \\[3mm]
The air removed from the coolant water degassing system and the biological shield area passes a noble gas retention system before being diluted by non-contaminated air at a rate of 166,000 m$^{3}$ h$^{-1}$ and emitted into the atmosphere. Data are not available on the system's retention times for argon, but using its value of 60 h for krypton and the argon to krypton ratios of activated carbon published (Bazan et al., 2011) and measured by ourselves (Heise, 2019) indicate argon retention times less than 24 hours. In the following, these are always set to 12 hours. \\[5mm] 
{\bf 2.5 Neutronic reactor simulations} \\[3mm]
For the in-core pathways, production rates and fuel inventories of $^{37}$Ar were calculated using the TRITON sequence (DeHart and Bowman, 2011) of the SCALE code system (Rearden and Jessee, 2016). Assuming vertically constant neutron flux densities over the active length of the fuel elements (3.9 m), their values along the radial cross section depicted in Fig. 2 were estimated using the 1-D discrete ordinates code XSDRN of SCALE. All simulations employed the ENDF/B-VII nuclear data libraries of SCALE (Ilas et al., 2012). \\[5mm] 
{\bf 2.6 Atmospheric dispersion} \\[3mm]
Transport and dilution of the calculated $^{37}$Ar emissions within the atmosphere were estimated for checking, whether pressurized water reactors may adversely affect the detectability of radioargon signals originating from nuclear weapon testing. This was performed with the generic dilution factors derived by Eslinger et al. (2015) for long distance (100 - 3.000 km) transport of noble gases. These were derived from an ensemble of 3650 simulations, which employed a Lagrangian model for calculating the optimum detector position in space and time if a 1 hour release was assumed at 10 release locations (both on the northern and southern hemisphere) and meteorological data for each day of the year 2011 (Eslinger et al., 2015). For routine emissions, these dilution factors may underestimate concentrations, as they do not take the superposition of consecutive 1 h release puffs into account. We considered this by using the upper 90 \% bound of the dilution factors given by Eslinger et al. (2015) instead of their median. For the peak emissions considered in this study, their highest 24 h mean has been assumed as source term in order to take a 24 h sampling time into account for the $^{37}$Ar analyses. \\[9mm]
         %
         %
\large
{\sf 3. RESULTS AND DISCUSSION} \\[3mm]
\normalsize
In this section, estimated emission rates are presented individually for each of the four potential pathways identified, followed by an evaluation of total emissions comparing our results to measurements and taking into account atmospheric transport and dilution. \\[5mm]
{\bf 3.1 Pathway (1): Activation of argon in cooling water} \\[3mm]
Estimated atmospheric emissions of $^{37}$Ar with time after the primary cooling pumps and the degassing system(s) are activated during the reactor restart operations are shown in Fig. 3. Due to the time required for the reactor to become critical and to increase its power level (see Fig. 1) emissions start to rise after ca. 1 day, reach a maximum speedily and decline to close to zero within 5 days. This dynamics reflects that the increase of $^{37}$Ar production rates with power become exceeded by the fast and effective reduction of its precursor $^{36}$Ar caused by the coolant water degassing. This effect is also documented by the decrease of peak emissions by almost an order of magnitude if the second system is operating as well. \\[2mm]
Usually after a refueling shutdown there are physics tests for checking that the neutronic operation characteristics of the modified core is as predicted and in accordance with the safety limits. These will cause a delay in increasing the power level, which due to the degassing will result in smaller $^{37}$Ar production rates compared to those shown in Fig. 3. Thus, our results represent a conservative estimate of potential $^{37}$Ar emissions by this pathway. \\[2mm]
If the reactor is restarted inadvertantly without any of the degassing system operating, the produced $^{37}$Ar will accumulate within the primary cooling system. Its emissions after both systems are activated are shown in Fig. 4 for various prior operation times. After the 12 hour delay caused by the noble gas retention systems of the plant, there is a peak emission of $^{37}$Ar. Due to the effective removal of all argon isotopes by the degassing systems emissions drop off effectively within two days, reflecting the degassing systems' removal rate. For short times after restart peak values increase linearly with the accumulation period, but with increasing time they level off due to the 37.5 day half-life of $^{37}$Ar. \\[2mm]
Replacing 10 m$^{3}$ d$^{-1}$ of the primary cooling water for limiting its tritium concentrations results in an $^{37}$Ar atmospheric emission of 14.1 Bq m$^{-3}$ at equilibrium due to activitation of the residual $^{36}$Ar  present in the degassed makeup water. \\[5mm]
{\bf 3.2 Pathway (2): Activation of calcium in fuel} \\[3mm]
If a major defect of one fuel rod present in the core develops after its $^{37}$Ar inventory has attained equilibrium, estimated emission rates both during rated power operation and caused by a shutdown transient are shown in Fig. 5. Although marked peaks are apparent from release of the $^{37}$Ar accumulated in the fuel rod gap and plenum, both when the defect develops and after a power transient, their concentrations in stack emissions are negligibly small and do not produce any detectable $^{37}$Ar background within the atmosphere. This is a marked difference from xenon emissions of nuclear power reactors, which for its major radioactive isotope $^{133}$Xe mainly result from fuel rod defects (Lewis et al, 1997), notably after reactor shutdown when the noble gas retention systems are bypassed (Saey et al., 2013). This discrepancy results from the fact that the concentration of calcium as the precursor of $^{37}$Ar in the fuel is limited to trace levels (5 ppm) producing a small equilibrium activity of 1.7$\, \cdot \,$10$^{6}$ Bq kg$^{-1}$ compared to the high yield fission product $^{133}$Xe with an equilibrium inventory of 8.3$\, \cdot \,$10$^{13}$ Bq kg$^{-1}$.  \\[5mm]
{\bf 3.3 Out-of-core neutron flux densities} \\[3mm]
For estimating out-of-core radioargon production rates, knowledge of the residual neutron flux densities and energy distributions within the biological shield is required. Results are depicted in Fig. 6 along the active fuel zone for rated power operation (see Fig. 2). For clarity, the original 238 energy group flux densities used for all neutronic calculations are collapsed into three broad group cross sections which represent thermal, epithermal and fast neutron energies. The total neutron flux densities at the outer surface of the reactor pressure vessel are still above 10$^{9}$ cm$^{-2}$ s$^{-1}$, but are effectively moderated and absorbed by the two concrete sections of the biological shield. These results indicate that $^{37}$Ar production by activation of argon in the cavities and of calcium in the concrete of the biological shield are of potential significance.  \\[5mm]
{\bf 3.4 Pathway (3): Activation of air in the biological shield} \\[3mm]
At rated power, $^{37}$Ar production rates in the cavities of the biological shield are time-invariant. Contrary to the two in-core pathways considered previously, emissions of this radioisotope originating from these cavities are almost constant after it has attained its equilibrium concentration ca. 12 days after the reactor commenced operation at constant power. \\[2mm]
Fig. 7 depicts the $^{37}$Ar production rates within the two cavities due to activation of natural $^{36}$Ar within the neutron field of the active zone of the reactor core, which were calculated using the neutron flux densities shown in Fig. 6. As expected they follow the radial neutron flux density levels. With the air volumes present within the neutron flux field and the ventilation rates the $^{37}$Ar equilibrium concentrations given in Table 1 evolve. In the outer cavity these are almost three orders of magnitude lower than in the inner one, which reflects both the smaller neutron flux densities and the large dilution volume of the contaminated air ventilation and exhaust system. Only 6.1 \% of this activity is produced by activation of the $^{36}$Ar being present in the outer cavity, the remainder originates from leakage from the inner into the outer cavity. \\[5mm]
{\bf 3.5 Pathway (4): Activation of calcium in concrete} \\[3mm]
Production rates of $^{37}$Ar within the biological shield at rated power are included in Fig. 7 both for low and high Ca concrete. As these are almost constant within the first 10 cm of the inner cavity's concrete wall, their averages have been used with eqn. (21) for calculating radioargon diffusion fluxes into the inner cavity across the surface of the activation zone for both concrete types. \\[2mm]
Production rates within the $^{37}$Ar diffusion length in the concrete walls of the outer cavity decrease with distance from the core (Fig. 7). As to our best knowledge no analytical solution of eqn. (19) is available for such a production rate profile, we approximated these by their mean value within the outer 20 cm of concrete. Using eqn. (22) and neglecting the small contributions from $^{37}$Ar being produced in the outer concrete wall, its rated power equilibrium concentrations caused by this pathway in the outer cavity are given in Table 1. \\[2mm]
Activity concentrations built up by this pathway are more than three orders of magnitude higher in the inner than in the outer cavity. This again reflects the radial decrease of neutron flux and the higher dilution volume of the contaminated air ventilation system of the outer cavity. As for the activation of $^{36}$Ar in air, most of the $^{37}$Ar accumulating in the outer cavity originates from the inner cavity. As Table 1 reveals, contributions by the out-of-core nuclear pathways are in the same order of magnitude, their ratios depending on the calcium content of the concrete. \\[5mm]
{\bf 3.6 Ratios of $^{41}$Ar to $^{37}$Ar} \\[3mm]
In most countries, stack emissions of $^{41}$Ar are continuously monitored. Thus, it would be attractive if the emitted air shows an almost constant $^{41}$Ar:$^{37}$Ar activity ratio, from which the unknown $^{37}$Ar emissions could be easily deduced (Fay and Biegalski, 2013; Johnson et al., 2017). However, our simulations indicate that for pressurized water reactors such a constant ratio can not be expected. First, $^{41}$Ar originating from in-core production pathways (i.e. pathway (1) above) will pass the stack noble gas retention systems and because of its short half-life (1.83 h) will largely decay before being emitted. Second, $^{41}$Ar built up by activation of air within the cavities of the biological shield will closely follow routine short-term variations of reactor power and of the contaminated air exhaust system operation mode. \\[5mm]
{\bf 3.7 Emissions and atmospheric concentrations} \\[3mm]
The in-core argon activation pathways evaluated above result in transient peak emissions, whereas the out-of-core pathways cause quite constant $^{37}$Ar emissions. Their estimated concentrations are given in Table 2, which result from taking the dilution of the outer cavity air by uncontaminated air from the reactor building into account. It should be noted that these data may slightly overestimate actual emissions, as they assume that after an extended period of rated power operation $^{37}$Ar concentrations have reached equilibrium. For comparison, Table 2 includes activity concentrations measured in the emissions of the German nuclear power plant Philippsburg-2 taken in December 2016 and September 2017 (Purtschert et al., in preparation), which are in good to excellent agreement with our predictions. \\[3mm]
Anthropogenic atmospheric concentrations of $^{37}$Ar calculated from routine emissions are shown in Fig. 8 for its various production pathways. Even at short distances from the emitting reactor, their concentrations become by several orders of magnitude smaller than the range of cosmogenic background concentrations specified by Riedmann and Purtschert (2011), which have been measured in air in Bern (Switzerland) during the last decades. \\[3mm]
 In Fig. 9 the $^{37}$Ar concentrations in air are shown, which result from our calculated peak 24 h  emissions after a restart following a reactor shutdown and contact of the primary cooling water with air. At distances of more than 100 km from the emitter, cosmogenic concentrations of $^{37}$Ar in the atmosphere will be exceeded only if the reactor's purification and degassing system is not operating for a prolonged period of time, allowing $^{37}$Ar to accumulate, and is switched on afterwards. However, such an operation mode constitutes a significant violation of the operation regulations and thus will occur rather infrequently. It will also cause elevated emissions of $^{41}$Ar accumulated in the primary cooling water. This short-lived isotope attains its saturation concentration within a few hours with maximum 24 h mean emissions of 6$\, \cdot \,$10$^{8}$ Bq m$^{-3}$ after the purification system is started. These are by several  orders of magnitude higher than the $^{37}$Ar emissions caused by such an event and thus provide a convenient option for discriminating such transient signals from $^{37}$Ar emitted after underground nuclear explosions. \\[9mm]
\large
{\sf 4. CONCLUSIONS} \\[3mm]
\normalsize
We have evaluated the pathways and atmospheric source terms of $^{37}$Ar in pressurised water reactors, as these constitute the by far most frequent commercial nuclear reactor type. It is produced via $^{36}$Ar$\,$(n,$\gamma$)$ \, ^{37}$Ar and $^{40}$Ca$\,$(n,$\alpha$)$ \, ^{37}$Ar reactions both in and out of the reactor core. \\[2mm]
In core, $^{37}$Ar emissions originate from activation of $^{36}$Ar at reactor restart after shutdown events with access of the primary coolant to air, in case of fuel rod defects from activation of calcium impurities of the uranium oxide, and from activation of residual $^{36}$Ar in makeup water during routine operation. The first two processes result in transient $^{37}$Ar source terms. The $^{37}$Ar signal caused by a fuel rod defect is negligible. After reactor restart, $^{37}$Ar concentrations in air become small compared to its cosmogenic background already at distances well below 100 km from the emitting plant. This becomes different, if after reactor restart its primary cooling water degassing system is switched on only after prolonged operation time. However, such an event also causes high emissions of $^{41}$Ar, which can be used to distinguish it from $^{37}$Ar originating from an underground nuclear explosion. Emissions caused by the third in core production process are almost constant, but small and become negligible at distances well below 100 km from the plant. \\[2mm]
Out of core, $^{37}$Ar results from activation of $^{36}$Ar in air and of $^{40}$Ca in concrete of the biological shield. Emissions due to these processes are almost constant, but small and below cosmogenic levels of $^{37}$Ar at distances well below 100 km from the plant. Our calculated source terms agree well with $^{37}$Ar measured in samples taken at the stack of a pressurised water reactor (Purtschert et al., in preparation). \\[2mm]
In summary, our study underlines the attractiveness of $^{37}$Ar as a potential indicator of underground nuclear explosions for complementing the radioxenon isotopes of the Comprehensive Nuclear-Test-Ban Treaty (CTBT) verification regime. \\[9mm]
{\bf Acknowledgements} \\[2mm]
 Financial support of this research provided by the German Foundation for Peace Research is gratefully acknowledged. We are grateful to Andreas Schmieder, Framatome Advanced Nuclear Fuels GmbH, for providing data on the calcium content of UO$_{2}$ fuel, to Gernot Kirchner, LafargeHolzim, for his information on concrete production and composition, and to Gerd B\"ohm, Andreas Bollh\"ofer, Andreas Deller, Ingeborg Krol, Christopher Strobl, Clemens Schlosser (all German Federal Office for Radiation Protection) and Roland Purtschert (University of Bern) for their support and many stimulating discussions. \\[9mm]
         %
         %
\large
{\sf REFERENCES} \\[5mm]
\normalsize
Aalseth, C.E., Day, A.R., Haas, D.A., Hoppe, E.W., Hyronimus, B.J., Keillor, M.E., Mace, E.K., Orrell, J.L., Seifert, A., Woods, V.T., 2011. Measurement of $^{37}$Ar to support technology for On-site Inspection under the Comprehensive Nuclear Test Ban Treaty. Nucl. Instr. Meth. Phys. Res. A652, 58-61. https://doi.org/10.1016/j.nima.2010.09.135. \\[3mm]
Bazan, R.E., Bastos-Neto, M., Moeller, A., Dreisbach, F., Staudt, R., 2011. Adsorption equilibria of O$_{2}$, Ar, Kr and Xe on activated carbon and zeolites: single component and mixture data. Adsorption 17, 371-383. https://doi.org/10.1007/s10450-011-9337-3. \\[3mm]
Carrigan, C.R., Heinle, R.A., Hudson, G.B., Nitao, J.J., Zucca, J.J., 1996. Trace gas emissions on geological faults as indicators of underground nuclear testing. Nature 382, 528-531. \\[3mm]
DeHart, M.D., Bowman, S.M., 2011. Reactor Physics Methods and Capabilities in SCALE. Nucl. Technol. 174, 196-213. https://doi.org/10.13182/NT174-196. \\[3mm]
D\"orr, H., M\"unnich, K.O., 1990. $^{222}$Rn flux and soil air concentration profiles in West-Germany. Soil $^{222}$Rn as tracer for gas transport in the unsaturated soil zone. Tellus B 42, 20-28.  https://doi.org/10.1034/j.1600-0889.1990.t01-1-00003.x \\[3mm]
Eslinger, P.W., Boywer, T.D., Cameron, I.M., Hayes, J.C., Miley, H.S., 2015. Atmospheric plume progression as a function of time and distance from the release point for radioactive isotopes. J. Environ. Radioact. 148, 123-129. https://doi.org/10.1016/j. jenvrad.2015.06.022. \\[3mm]
Fay, A.G., Biegalski, S.R., 2013. Contributions to the $^{37}$Ar background by research reactor operation. J. Radioanal. Nucl. Chem. 296, 273-277. https://doi.org/10.1007/s10967-012-1968-7. \\[3mm]
Haas, D.A., Miley, H.S., Orrell, J.L., Aalseth, C.E., Bowyer, T.W., Hayes, J.C., McIntyre, J.I., 2010. The Science Case for $^{37}$Ar as a Monitor for Underground Nuclear Explosions. Pacific Northwest National Laboratory, Richland, PNNL-19458.
\\[3mm]
Heise, A., 2019. Machbarkeitsstudie zur Nutzung des Radioisotopes Argon-37 im Rahmen des Verifikationsregimes des
Umfassenden Kernwaffenteststopp-Vertrags. PhD Dissertation, Universit\"at Hamburg, Dept. of Physics, Hamburg (in German). \\[3mm]
Hering, W., 1983. The KWU fission gas release model for LWR fuel rods. J. Nuc. Mater. 114, 41-49. https://doi.org/10.1016/0022-3115(83)90071-5. \\[3mm]
Ilas, G., Gauld, I.C., Radulescu, C., 2012. Validation of New Depletion Capabilities and ENDF/B-VII data libraries in SCALE. Ann. Nucl. Energy 46, 43-55. https://dx.doi.org/\-10.1016/j.\-anucene.2012.03.012. \\[3mm]
International Atomic Energy Agency, 2010. Review of Fuel Failures in Water Cooled Reactors. IAEA Nuclear Energy Series No. NF-T-2.1, IAEA, Vienna. \\[3mm]
International Atomic Energy Agency, 2019. Nuclear Power Reactors in the World. 2019 Edition. IAEA-RDS-2/39, IAEA, Vienna. \\[3mm]
International Union of Pure and Applied Chemistry, 1980. Solubility Data Series, Volume 4 -- Argon. Pergamon Press, Oxford. \\[3mm]
Johnson, C., Biegalski, S.R., Artnak, E.J., Moll, E., Haas, D.A., Lowrey, J.D., Aalseth, C.E., Seifert, A., Mace, E.K., Woods, V.T., Humble, P., 2017. Production and release rate of $^{37}$Ar from the UT TRIGA Mark-II research reactor.  J. Environ. Radioact. 167, 249-253. https://doi.org/10.1016/j.jenvrad.2016.11.017. \\[3mm]
Kalinowski, M.B., Pistner, C., 2013. Isotopic signature of atmospheric xenon released from light water reactors. J. Environ. Radioact. 88, 215-235. https://doi.org/10.1016/j. jenvrad.2006.02.003. \\[3mm]
Lewis, B.J., Iglesias, F.C., Postma, A.K., Steininger, D.A., 1997. Iodine spiking model for pressurized water reactors. J. Nucl. Mater. 244, 153-167. https://doi.org/ 10.1016/\-S0022-3115(96)00723-4. \\[3mm]
Lewis, B.J., Chan, P.K., El-Jaby, A., Iglesias, F.C., Fitchett, A., 2017. Fission product release modelling for application of fuel-failure monitoring and detection -- An overview. J. Nuc. Mater. 489, 64-83. https://dx.doi.org/10.1016/j.nucmat.2017.03.037. \\[3mm]
Manzel, R., Walker, C.T., 2002. EPMA and SEM of fuel samples from PWR rods with an average burnup of around 100 MWD/kgHM. J. Nucl. Mater. 301, 170-182. https://doi.org/10.1016/S0022-3115(01)00753-X. \\[3mm]
%
%
Rearden, B.T., Jessee, M.A. (Eds.), 2016. SCALE Code System. ORNL/TM-2005/39, Version 6.2.1, Oak Ridge National Laboratory, Oak Ridge, TN. Available from Radiation Safety Information Computational Center as CCC-834. \\[3mm]
Rest, J., Cooper, M.W.D., Spino, J., Turnbull, J.A., Van Uffelen, P., Walker, C.T., 2019. Fission gas release from UO$_{2}$ nuclear fuel: A review. J. Nucl. Mater. 513, 310-345. https://doi.org/10.1016/j.jnucmat.2018.08.019. \\[3mm]
Riedmann, R.A., Purtschert, R., 2011. Natural $^{37}$Ar concentrations in soil air: implications for monitoring underground nuclear explosions. Environ. Sci. Technol. 45, 8656-8664. https://pubs.acs.org/doi/10.1021/es201192u. \\[3mm]
Saey, P.R.J., 2009. The influence of radiopharmaceutical isotope production on the global radioxenon background. J. Environ. Radioact. 100, 396-406. https://doi.org/10.1016/j. jenvrad.2009.01.004. \\[3mm]
Saey, P.R.J., Ringbom, A., Bowyer, T.W., Z\"ahringer, M., Auer, M., Faanhof, A., Labusch\-agne, C., Al-Rashidi, M.S., Tippawan, U., Verboomen, B., 2013. Worldwide measurements of radioxenon background near isotope production facilities, a nuclear power plant and at remote sites: the ''EU/JA-II'' Project. J. Radioanal. Nucl. Chem. 296, 1133-2142. DOI 10.1007/s10967-012-2025-2. \\[3mm]
Sahoo, B.K., Sapra, B.K., Gaware, J.J, Kanse, S.D., Mayya, Y.S., 2011. A model to predict radon exhalation from walls to indoor air based on the exhalation from building material samples. Sci. Total Environ. 409, 2635-2641. https://doi.org/10.1016/j.scitotenv. 2011.03.031. \\[3mm]
Shen, L., Chen, Z., 2007. Critical review of the impact of tortuosity on diffusion. Chem. Eng. Sci. 62, 3748-3755. https://doi.org/10.1016/j.ces.2018.04.074 
%
\newpage
{\bf Table 1:} Estimated equilibrium concentrations of $^{37}$Ar in the air cavities of the biological shield resulting from the various reaction pathways considered. \\[5mm]
{\bf Table 2:}  Routine emissions of $^{37}$Ar caused by the out-of-core activation pathways calculated here and comparison with measurements in exhaust air of the German Philippsburg-2 nuclear power plant (Purtschert et al., in preparation).
\newpage
{\bf Table 1} \\[1cm]
\renewcommand{\baselinestretch}{1}
\begin{center}
\begin{tabular}{c|c|cc} \hhline{*{4}{-}} \rule[-0mm]{0mm}{6mm}Activation & Cavity$^{(a)}$ & \multicolumn{2}{c}{$^{37}$Ar concentration [Bq m$^{-3}$]} \\
  route &  & \multicolumn{2}{c}{for concrete with} \\
  &  &  8.7 \% Ca  &  37.5 \% Ca \\    \hhline{*{4}{-}}
 $^{36}$Ar\,(n,$\gamma$)\,$^{37}$Ar \rule[-0mm]{0mm}{6mm} &  inner  &  \multicolumn{2}{c}{3.98$\, \cdot  \, $10$^{6}$} \\
     &  outer  &  \multicolumn{2}{c}{2.50$\, \cdot  \, $10$^{2}$} \\
	 &   \rule[-4mm]{0mm}{4mm}outer, leakage$^{(b)}$  &  \multicolumn{2}{c}{3.82$\, \cdot  \, $10$^{3}$}  \\
 $^{40}$Ca\,(n,$\alpha$)\,$^{37}$Ar  &  inner  & 1.81$\, \cdot  \, $10$^{6}$  & 7.79$\, \cdot  \, $10$^{6}$  \\
     &  outer  & 2.67$\, \cdot \, $10$^{2}$  &  1.15$\, \cdot \, $10$^{3}$  \\
		 &  \rule[-4mm]{0mm}{4mm}outer, leakage$^{(b)}$     & 1.73$\, \cdot \, $10$^{3}$  &  7.46$\, \cdot \, 
		$10$^{3}$  \\
		total  &  inner  &   5.79$\, \cdot  \, $10$^{6}$  & 1.18$\, \cdot  \, $10$^{7}$  \\
		 &  outer  &  6.07$\, \cdot  \, $10$^{3}$  & 1.27$\, \cdot  \, $10$^{4}$  \\[1mm] \hhline{*{4}{-}}
\end{tabular} \\[3mm]
\end{center}
\makebox[15mm]{}$^{(a)}$ the outer cavity volume includes the contaminated air exhaust system \\
\newpage
{\bf Table 2} \\[1cm]
\renewcommand{\baselinestretch}{1}
\begin{center}
\begin{tabular}{d|cc}  \hhline{*{3}{-}}  \multicolumn{1}{c|}{Ca in concrete} &   \multicolumn{2}{c}{\rule[-0mm]{0mm}{6mm}$^{37}$Ar concentration [Bq m$^{-3}$]} \\
    \rule[-0mm]{-1mm}{3mm}  [\%] &    calculated  &  measured  $^{(a)}$\\[1mm]       \hhline{*{3}{-}}
   8.7   &   \rule[-0mm]{0mm}{6mm}36.5 &     21.7$\: \pm \:$0.2   \\
	          &   &  24.9$\: \pm \:$0.7   \\
						&   &  10.2$\: \pm \:$0.3     \\[1mm]
 37.5   &  76.4    \\[1mm] \hhline{*{3}{-}}
\end{tabular} \\[3mm]
\end{center}
\makebox[15mm]{}$^{(a)}$ Purtschert et al., in preparation 
\newpage
{\bf Fig. 1:} Power-time diagram of a fast cold start of a PWR after a refueling and maintenance shutdown with 1: start of degassing system(s), 2: start of primary cooling pumps, 3: criticality of core. \\[5mm]
{\bf Fig. 2:} Schematic cross section (not to scale) of reactor core and biological shield with 1: reactor pressure vessel, 2: inner cavity, 3: inner shield, 4: outer cavity, 5: outer shield. The marked vertical section denotes the high neutron flux region during reactor operation. \\[5mm]
{\bf Fig. 3:} $^{37}$Ar emissions after reactor start from activation of argon dissolved in the primary cooling water with two (bold line) and a single degassing system working (dotted line), respectively. \\[5mm]
{\bf Fig. 4:} $^{37}$Ar emissions after start and full power operation for 1 day (dotted), 5 days (dashed) and 50 days (bold) with degassing systems switched on only then (set to time zero). \\[5mm]
{\bf Fig. 5:} $^{37}$Ar emissions after a fuel rod defect developed at rated power and after a reactor shutdown transient 7 days afterwards with two (bold line) and a single degassing system working (dotted line), respectively. \\[5mm]
{\bf Fig. 6:} Radial neutron flux densities along the reactor core edge and the biological shield for thermal (10$^{-5}\,-\,$0.4 eV, dashed line), epithermal (0.4$\,-\,$10$^{6}$ eV, solid line) and fast neutrons (10$^{6}\,-\,$2$\cdot$10$^{7}$ eV, dot-dashed line); numbers denote cooling water (1),  steel (2), reactor pressure vessel (3), air filled cavity (4) and concrete (5). \\[5mm]
{\bf Fig. 7:} $^{37}$Ar production rates within cavities and concrete (dashed line: 8.7 \% Ca, solid line: 37.5 \% Ca) of the biological shield as a function of radial distance from the reactor core; for the meaning of the numbers see Fig. 6. \\[5mm]
{\bf Fig. 8:} Atmospheric concentrations of anthropogenic $^{37}$Ar resulting from routine emissions by in-core activation of residual $^{36}$Ar in makeup water (dashed line)  and by out-of-core pathways in pressurised water reactors with 8.7 \% Ca (dotted line) and 37.5 \% Ca (solid line) in concrete; the shaded area denotes its cosmogenic background (Riedmann and Purtschert, 2011). \\[5mm]
{\bf Fig. 9:} Atmospheric concentrations of anthropogenic $^{37}$Ar resulting from transient emissions of in-core pathways caused by reactor restart after opening of the reactor vessel with the coolant water purification system operating (dashed-dotted line) and not operating for 1 d (dotted line), 5 d (dashed line) and 50 d (solid line), respectively; the shaded area denotes its cosmogenic background (Riedmann and Purtschert, 2011). \\[5mm]
\newpage
\begin{figure}[t]
\makebox[16cm][l]{{\bf Fig. 1}} \\[3cm]
\begin{center}
\includegraphics[width=170mm]{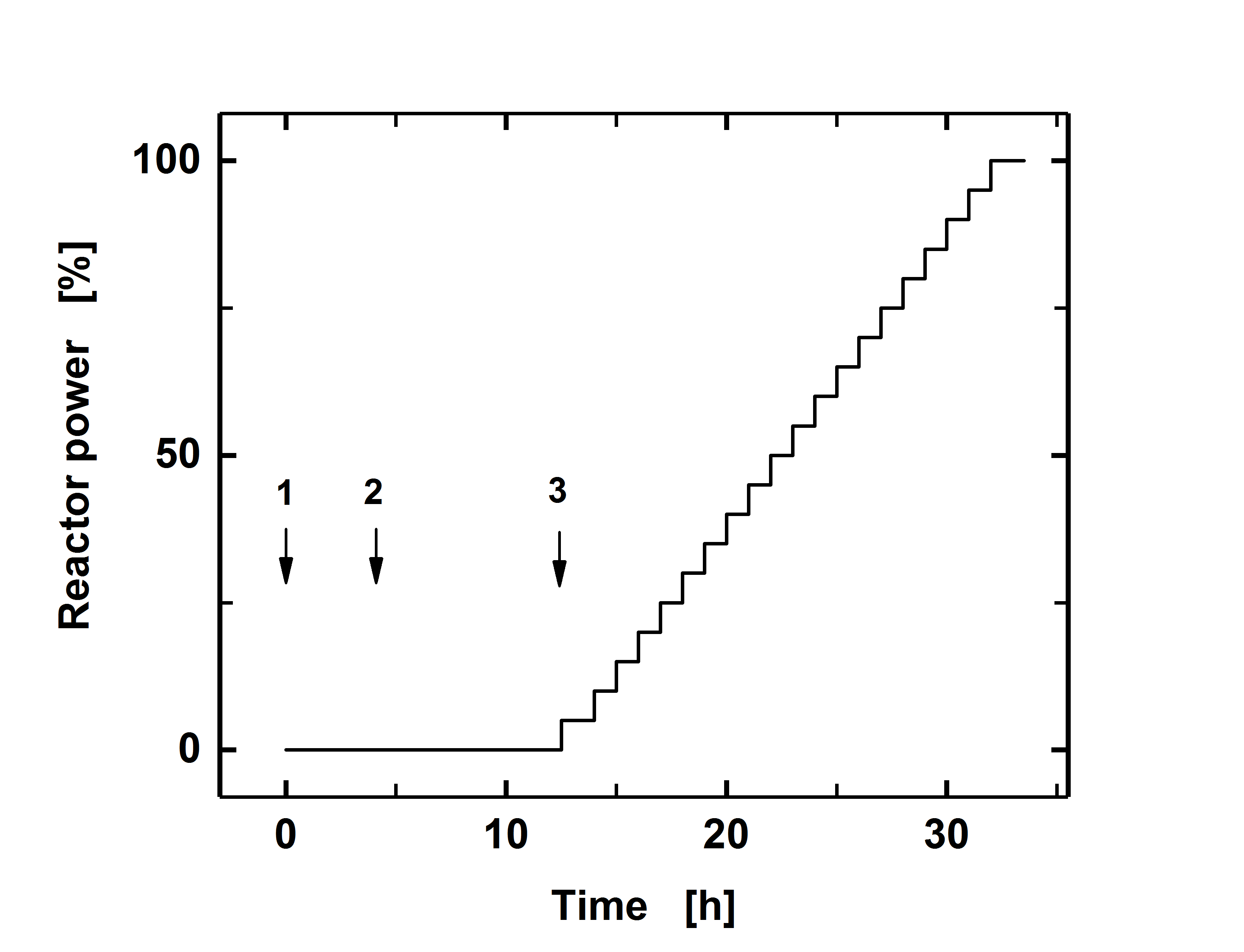}
\end{center}
\end{figure}
\newpage
\begin{figure}[t]
\makebox[16cm][l]{{\bf Fig. 2}} \\[3cm]
\begin{center}
\includegraphics[width=140mm]{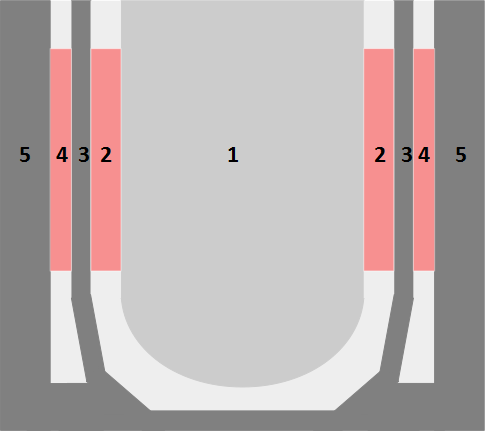}
\end{center}
\end{figure}
\newpage
\begin{figure}[t]
\makebox[16cm][l]{{\bf Fig. 3}} \\[3cm]
\begin{center}
\includegraphics[width=170mm]{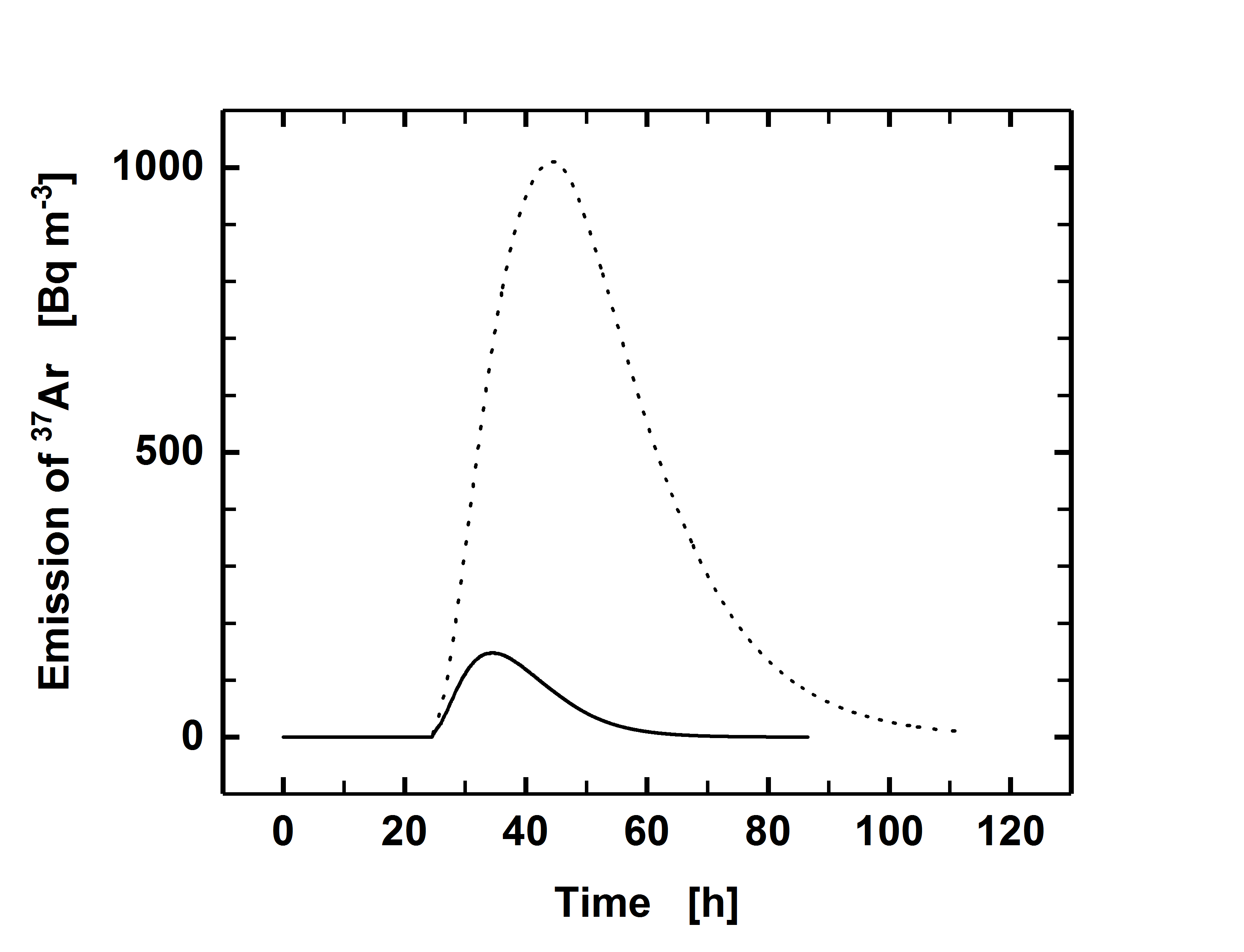}
\end{center}
\end{figure}
\newpage
\begin{figure}[t]
\makebox[16cm][l]{{\bf Fig. 4}} \\[3cm]
\begin{center}
\includegraphics[width=170mm]{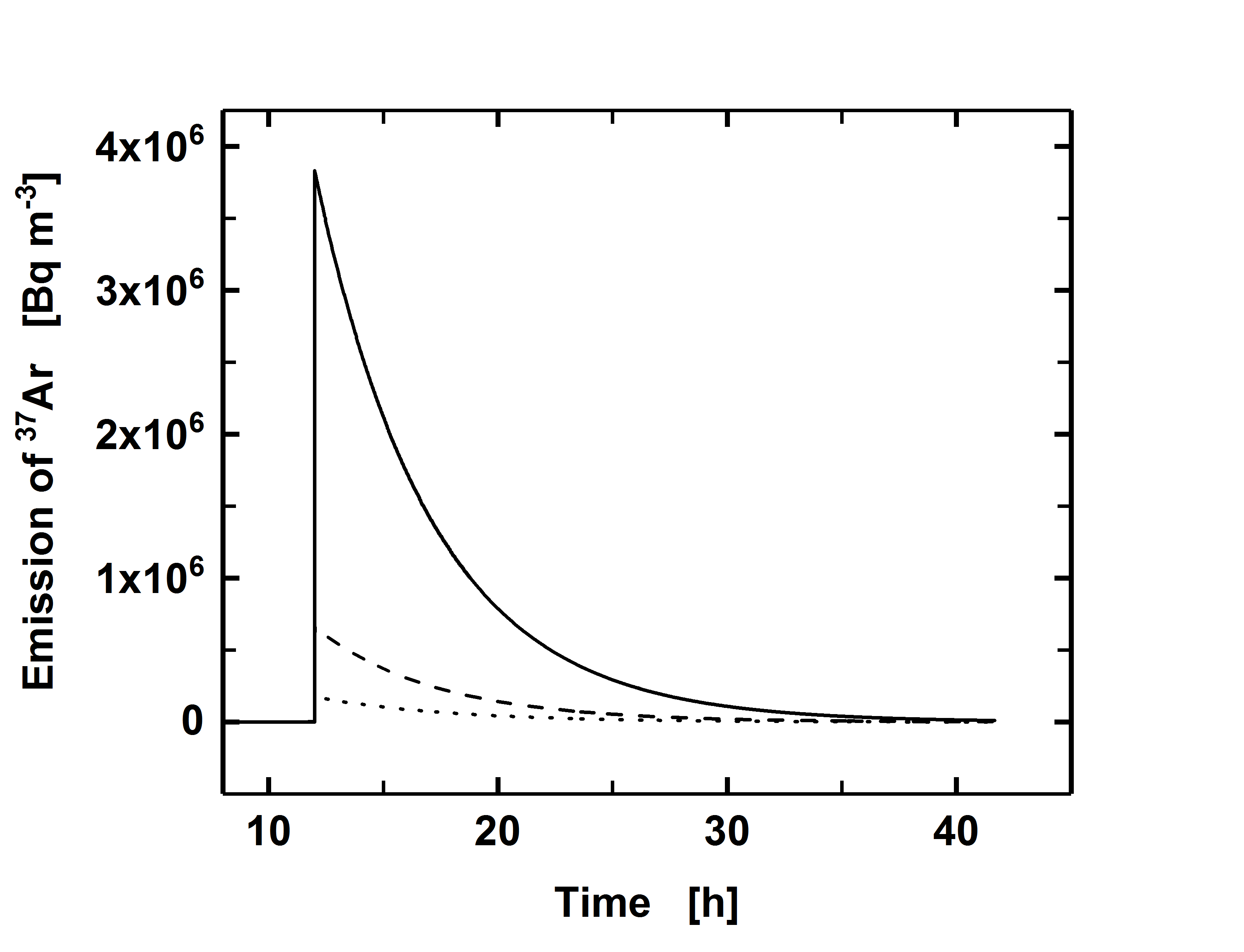}
\end{center}
\end{figure}
\newpage
\begin{figure}[t]
\makebox[16cm][l]{{\bf Fig. 5}} \\[3cm]
\begin{center}
\includegraphics[width=170mm]{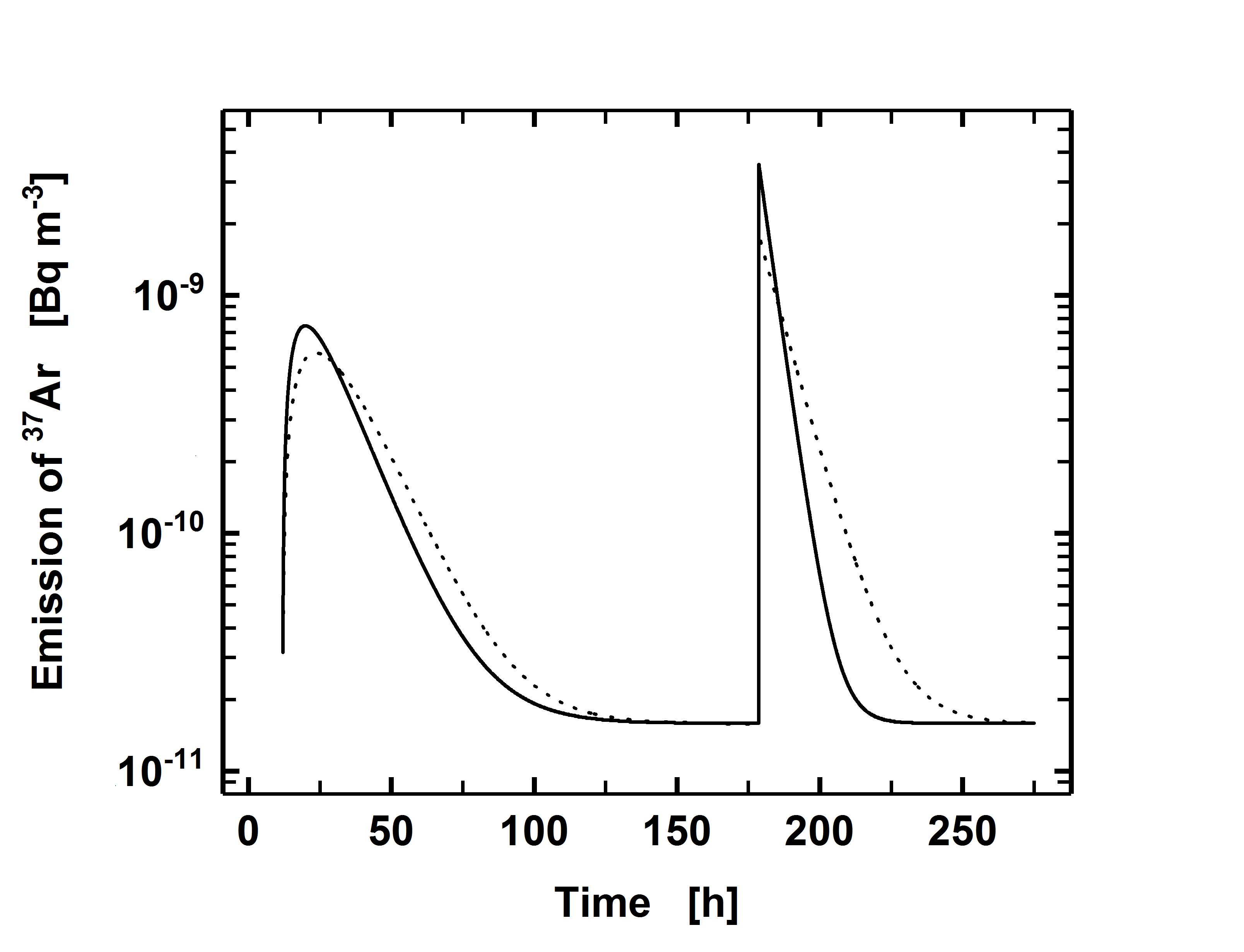}
\end{center}
\end{figure}
\newpage
\begin{figure}[t]
\makebox[16cm][l]{{\bf Fig. 6}} \\[3cm]
\begin{center}
\includegraphics[width=170mm]{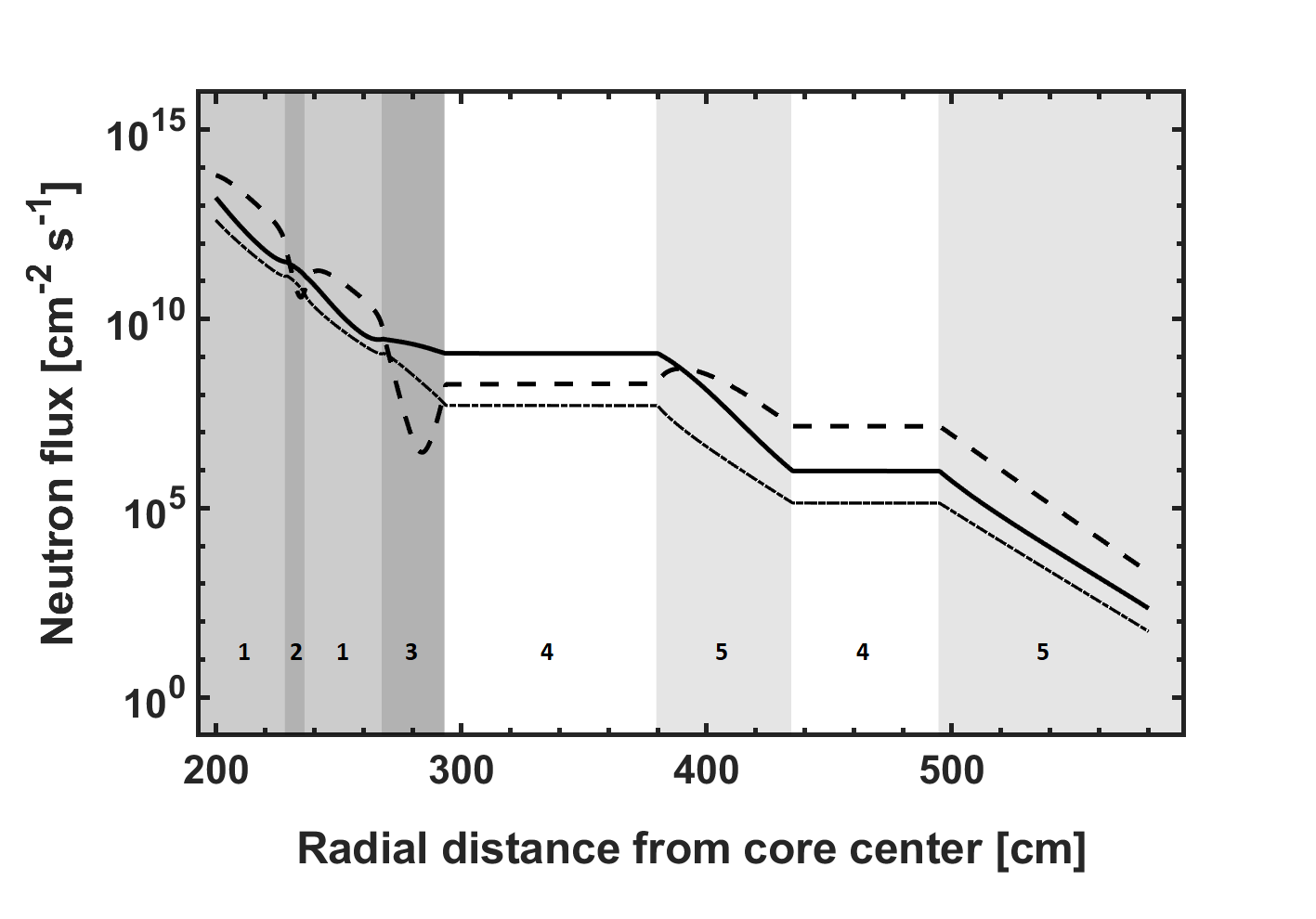}
\end{center}
\end{figure}
\newpage
\begin{figure}[t]
\makebox[16cm][l]{{\bf Fig. 7}} \\[3cm]
\begin{center}
\includegraphics[width=170mm]{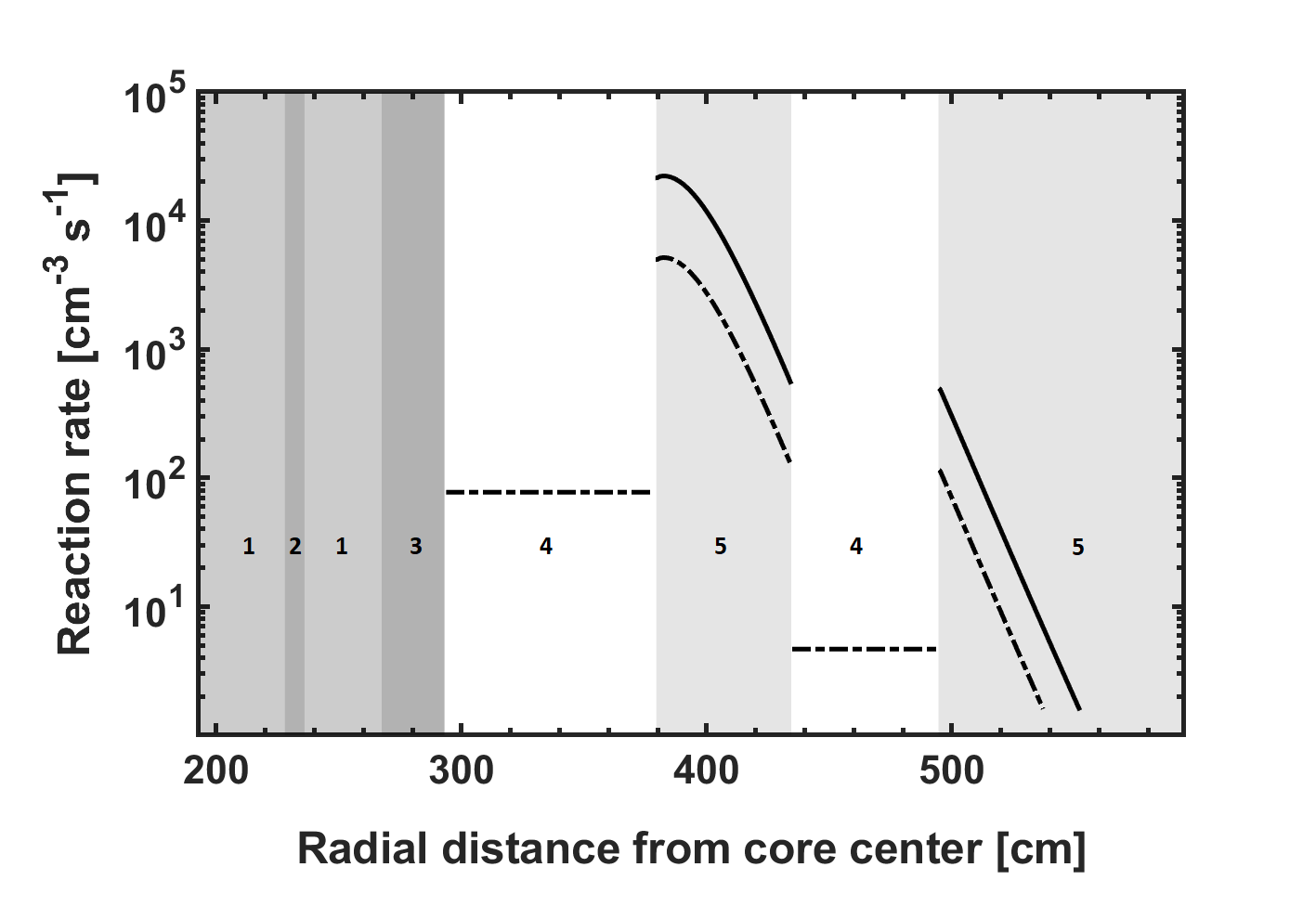}
\end{center}
\end{figure}
\newpage
\begin{figure}[t]
\makebox[16cm][l]{{\bf Fig. 8}} \\[3cm]
\begin{center}
\includegraphics[width=170mm]{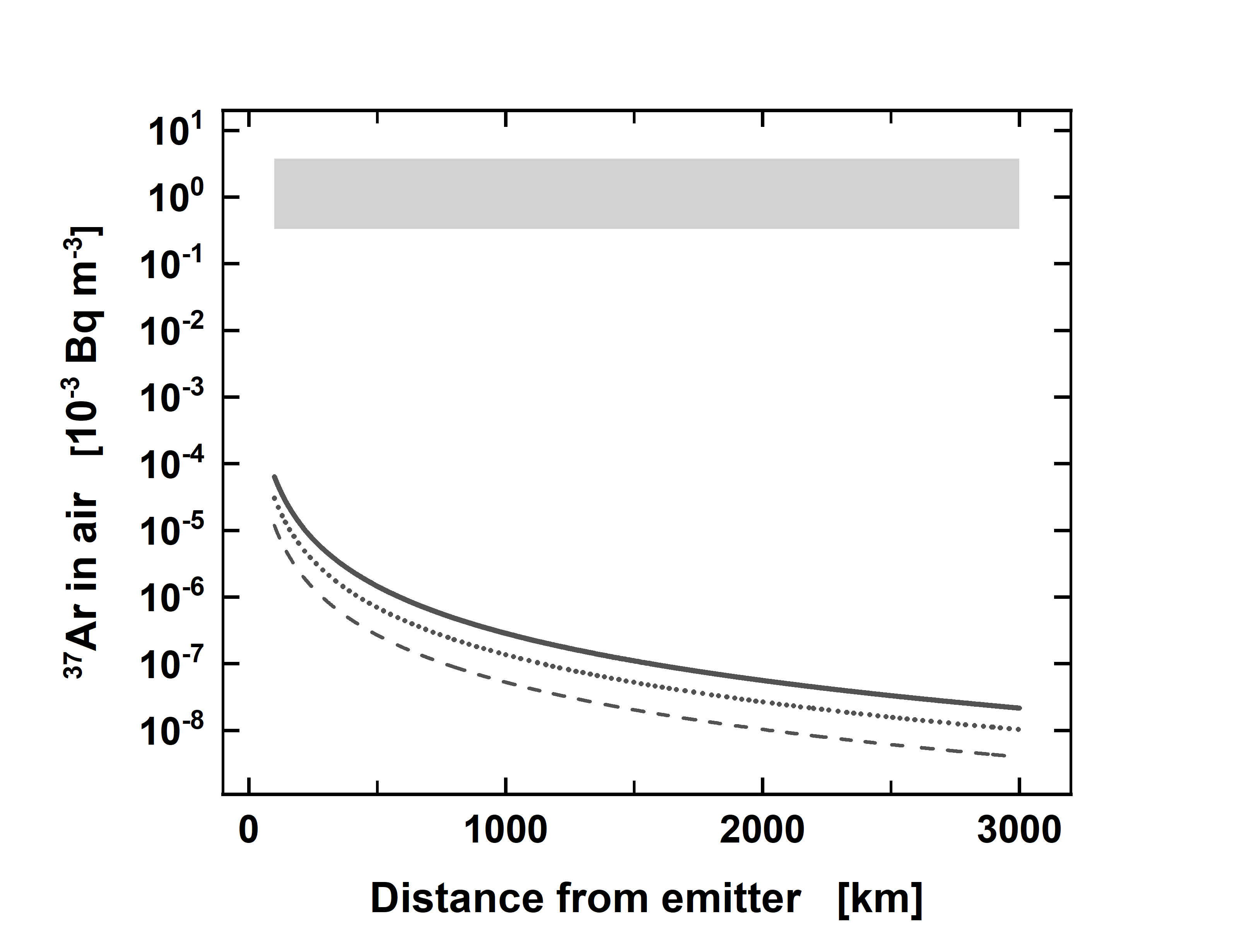}
\end{center}
\end{figure}
\newpage
\begin{figure}[t]
\makebox[16cm][l]{{\bf Fig. 9}} \\[3cm]
\begin{center}
\includegraphics[width=170mm]{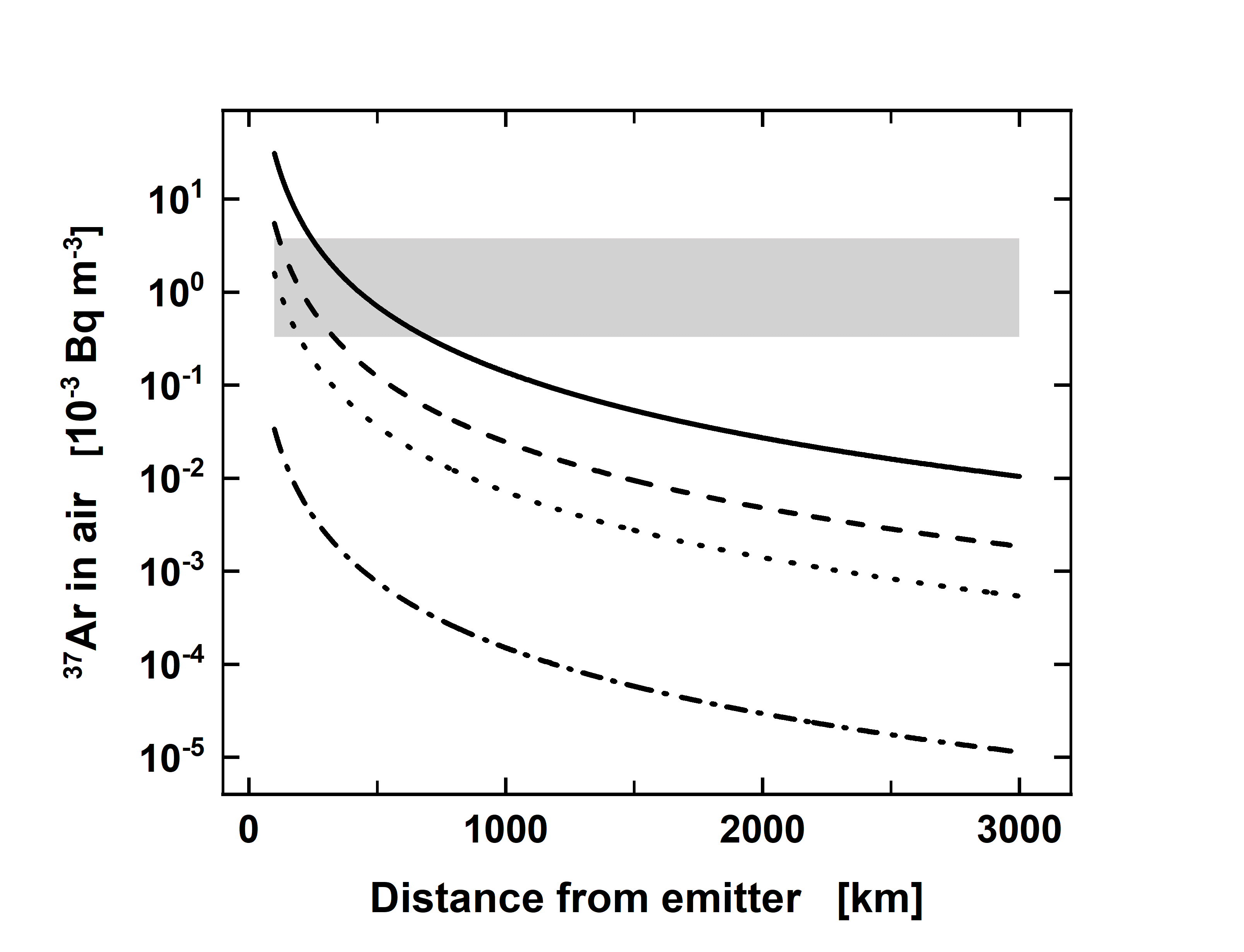}
\end{center}
\end{figure}
\end{document}